\documentclass[aps,preprint,onecolumn,eqsecnum]{revtex4}
\usepackage{dcolumn}
\usepackage{bm}
\usepackage{mathrsfs}
\usepackage{amssymb}
\usepackage{amsmath}
\usepackage{amsfonts}
\usepackage{color}
\usepackage{ulem}

\usepackage[]{graphicx}
\begin{document}
\newcommand{\half}{\frac{1}{2}}
\title{A causal and continuous interpretation of the quantum theory: About an original manuscript by David Bohm sent to Louis de Broglie in 1951}
\author{Aur\'elien Drezet$^{1}$}                  
\address{$^2$ Univ. Grenoble Alpes, CNRS, Grenoble INP, Institut Neel, F-38000 Grenoble, France}
\email{aurelien.drezet@neel.cnrs.fr}
\author{Benjamin Stock$^{2}$, }  
\address{$^2$ 1, Le Ch\^{e}ne Br\^{u}l\'{e}, 28400 Arcisses, France}

\begin{abstract}
The aim of this article is reproduce and analyze an original article of David Bohm sent to Louis de Broglie in 1951. This article is the older document of David Bohm about his well known hidden variable theory based on the pilot wave interpretation of Louis de Broglie.   We analyse the chronology and the history of this fascinating document. \\
  \begin{quote}\textit{ I have never been able to discover any well-founded reasons as to why there exists so high a degree of confidence in the general principles of the current form of the quantum theory} \cite{BohmImpl} p. 107.\end{quote}		
\end{abstract}

\maketitle
\section{General introduction}
The present article reproduces and studies an undated and apparently non analyzed manuscript  untitled `A causal and continuous interpretation of the quantum theory' sent to Louis de Broglie  by David Bohm in the year 1951 and concerning a pilot-wave quantum theory nowadays universally called de Broglie-Bohm interpretation or `Bohmian mechanics'. This text (referred here as ACCIQT) was found by us in the Archive Louis de Broglie at the French Academy of Science and has gone unnoticed until now \cite{Manuscript1951}.  From its content we believe this short text to be a primitive version of the two famous articles `A suggested interpretation of the quantum theory in terms of hidden variables. I and II' submitted together to the Physical Review  the 5$^{th}$ of July 1951 and published the 15$^{th}$ of January 1952 in the same volume \cite{Bohm1952a,Bohm1952b}.\\
The paper is organized as follows: First, in section \ref{ACCIQT} we reproduce the manuscript (a pdf version will be available XXXX). In section \ref{history} we discuss the chronology of this manuscript and in section \ref{section3} we analyse the physical contents and implications of ACCIQT for the understanding of Bohm work on hidden variables published in 1952 \cite{Bohm1952a,Bohm1952b}. Finally, we conclude by relating and connecting ACCIQT with the more general work  done by Bohm in order to understand and grasp the mystery of quantum mechanics         
\section{A Causal and Continuous Interpretation of the Quantum Theory: (ACCIQT)} \label{ACCIQT}

David Bohm \cite{Manuscript1951}\\ 
\textsl{Faculdade de Filosofia, Ciencias e Letras, Universidade de S\~{a}o Paulo, Sao Paulo, Brasil}\\

The usual interpretation of the quantum theory is based on the assumption that at the atomic level, the laws of nature are intrinsically statistical, in the sense that no more fundamental theory is possible that could causally and continuously account for quantum fluctuations in terms of at present "hidden" variables or parameters. This assumption leads to an extremely far-reaching change in our concepts concerning the nature of matter; for it requires us to renounce the possibility of even \underline{conceiving} in precise terms of the behavior of an individual system at a quantum level of accuracy \textbf{Ref.(1)}.	This change in our concepts is justified in part by its success in accounting for a very wide range of experimental phenomena, at least in the domain of distances larger than 10$^{-13}$ cm., and in part by the widely accepted belief that no
consistent causal and continuous interpretation of the mathematical equations of the quantum theory can possibly lead to the same results as those of the usual interpretation. In this note, however, the author wishes to call attention
to an alternative interpretation of the quantum theory, which leads to \underline{all}, of the experimental results predicted by the usual interpretation, but which permits us to retain the concept of a precise causal and continuous description of the motion of an \underline{individual} material system, even at a quantum level of accuracy. This interpretation, which is reported in 
 detail elsewhere \textbf{Ref.(2)} will be summarized here.\\
\indent We begin by regarding the electron as a particle having a precisely defined position and momentum at each instant of time. This particle is always accompanied, however, by a wave field, which we call the $\Psi$-field, and which satisfies Schr\"{o}dinger's equation. This field exerts a force on the particle, which is derivable from a ``quantum-mechanical'' potential,
\begin{eqnarray}
U=\frac{-\hbar^2}{2m}\frac{\boldsymbol{\nabla}^2|\Psi|}{|\Psi|}\label{eq1}
\end{eqnarray}
The equations of motion of the particle are then given by:
\begin{eqnarray}
m\frac{d^2}{dt^2}\textbf{x}=-\boldsymbol{\nabla}(U +V)\label{eq2}
\end{eqnarray}
where $V$ is the usual classical potential. It is the above ``quantum-mechanical'' potential that leads to all the characteristically new quantum-mechanical effects, and the classical limit is obtained when the effects of this potential on the particle can be neglected.\\
\indent The above equations of motion must, however, be supplemented by two mutually consistent assumptions. First, if we write
  $\Psi=R(\textbf{x})e^{iS(\textbf{x})/\hbar}$,  where $R$ and $S$ are real, then the momentum of the particle is restricted to $\mathbf{p}=\boldsymbol{\nabla}S(\textbf{x})$. This assumption is consistent in the sense that if it holds at a given time  then the equations of motion \ref{eq2} guarantee that it will hold for all time. Thus it it essentially a subsidiary condition.\\
\indent Secondly, we must assume that we do not \underline{in practice} predict or control the exact location of the particle, but that we 
have a statistical ensemble with a probability density of $P=|\Psi|^2$. The assumption of this ensemble is also consistent, in the sense that if it holds initially, then the equations of motion guarantee that it will hold for all time. The need for a statistical ensemble originates in the chaotic character of the particle motion (resembling Brownian motion), which arises whenever the particle interacts with other systems. Even if the initial position and momentum were known with perfect accuracy, the particle would soon diffuse over the entire region in which $|\Psi|$ was appreciable. Moreover, its motion would be so complicated that in practice we would 
be able to predict only the probability of a given location. 
It can be shown that after sufficient interactions have taken place, the probability density will tend to approach $P=|\Psi|^2$. 
Thus, the use of a statistical ensemble is here very similar to its use in classical statistical mechanics, where the statistical treatment is likewise made valid by the chaotic character of molecular motions brought about by collisions.\\
\indent It is shown that with these assumptions, we can describe 
in a causal\footnote{here Bohm wrote casual}  and continuous way \underline{all} of the results obtained  from the usual interpretation. For example, in the photo-electric effect, the transfer of a full quantum of energy to an atom in a short time by a very weak electromagnetic wave is 
made possible by the ``quantum-mechanical'' potential, which is not necessarily small when the wave amplitude is small (because this amplitude appears in the denominator of the potential). As a result, even in a wave of low amplitude, rapid and violent fluctuations of the potential occur, which can transfer a full quantum of energy in a time so short that the process appears in the usual observations to be discontinuous. The actual course of this transfer is in principle determined by the precise values of all the potentials at each point in space and time, but these potentials vary in such a chaotic way that in practice we can predict only the probability that a transfer will take place.\\
\indent As long as the equation governing the $\Psi$ field (Schr\"{o}dinger's equation) is linear and homogeneous, then our alternative interpretation leads to \underline{precisely} the same results for all experiments as are given by the usual interpretation. However, in our interpretation, it is consistent to contemplate modifications in the mathematical theory which could not consistently be made in the usual interpretation. For example, we can consider equations governing $\Psi$ which are non-linear and inhomogeneous, and which depend on the actual location of the particle. Moreover, the above modifications can be so chosen that they produce negligible changes in the atomic domain, where the present theory is known to be a good approximation, but significant changes in the domain of distances of the order of 10$^{-13}$cm, where present theories do not seem to be adequate. It is thus entirely possible that a correct theory of elementary particles will require the introduction of a causal and continuous interpretation of the quantum theory, such as the one described here.\\
\indent A theory of measurements has been developed, and it has been shown that as long as no mathematical modifications of the types described above are made, the uncertainty principle is obtained as a \underline{practical} limitation on the precision with which complementary variables such as position and momentum can be measured. However, if the mathematical formulation neeeds to be changed in almost any conceivable way in any domain (for example, at small distances), then measurements of unlimited precision can be shown to be possible in \underline{every} domain, including even those domains in which the present form of quantum theory is a good approximation.\\
\indent Our interpretation introduces what are essentially ``hidden'' causal parameters; namely, the precisely definable particle positions and momenta which do not appear ia the usual interpretation, but which along with the $\Psi$ field \textbf{Ref.(3)} determine in principle the \underline{actual} results of each \underline{individual} measurement process. At first sight, the existence of such ``hidden'' causal parameter would appear to contradict von Neumann's proof that no \underline{single} statistical distribution of ``hidden'' causal parameters could possibly account for all of the results of the usual interpretation of the quantum theory \textbf{Ref.(4)}. Von Neumann's proof is based, however, on the implicit assumption that the `hidden'' parameters are all in the observed system and not in the measuring apparatus. But in our interpretation, the ``hidden'' parameters are also in the measuring apparatus. Since different kinds of apparatus are needed in different kinds of measurements, the hidden parameters that determine the results of a momentum measurement differ from those which determine the results of a position measurement. Von Neumann's theorem therefore does not apply to our interpretation.\\
\indent After the work reported here had been completed, the author's attention was called to a similar interpretation suggested by de Broglie in 1926, but given up by him because of certain difficulties in the interpretation of a superposition of stationary state wave function \textbf{Ref.(5)}. De Broglie, however, did not carry his ideas to their logical conclusion \textbf{Ref.(6)}. The essential new steps needed were to improve the treatment of the problem of two bodies, and to construct a theory of measurements with the aid of the new interpretation. When these two steps are made then it can be shown that this interpretation leads to precisely the same results as are obtained with the usual interpretation.\\
\indent To sum up, we have developed a causal and continuous interpretation of the quantum theory. Such an interpretation is not only of general philosophical interest, but also has a possible significance for the development of new theories in those domains in which the present theory is inadequate.\\
\begin{quote}
\textbf{Ref.(1)} N. Bohr, Atomic Theory and Description of Nature. London: Cambridge University Press, 1934. \\
\textbf{Ref.(2)} D. Bohm, Phys. Rev. (to be published Dec. 15)\\
\textbf{Ref.(3)}) The $\Psi$ field can be defined with the aid of present types of measurements. Thus when the observable, $A$, is measured and found to have the eigenvalue, $a$, the wave function is known to be the corresponding eigenfunction $\Psi_a(\textbf{x})$ (except for a constant phase factor of no significance). \\
\textbf{Ref.(4)}) J. von Neumann, Mathematishe Grundlagen der Quantenmechanik. Berlin: Julius Springer, 1932.\\
\textbf{Ref.(5)}) L. de Broglie, Comptes Rendus de l'Acad\'{e}mie des sciences \textbf{183}, 447 (1926). Cf. also, Introduction \`{a} l'\'{e}tude de la M\'{e}canique Ondulatoire. Paris: Hermann, 1928 (English edition Methuen, London).\\
\textbf{Ref.(6)}) Cf. E. Madelung, Zeits. F. Physik, \textbf{40}, 322 (1927)\footnote{Bohm's reference was incorrectly written: E. Madelung, Zeits. F. Physik, \textbf{40}, 327 (1926).}. Madelung proposed a similar model of the quantum theory, but likewise did not carry it to its logical conclusion.
\end{quote} 
\section{A chronological inquiry concerning ACCIQT} \label{history}
 \indent The so called de Broglie-Bohm ontological or causal hidden-variable theory is nowadays accepted as an alternative description of quantum mechanics. The theory is empirically equivalent to standard quantum mechanics at least in the non-relativistic domain and opens new perspective for particle physics and cosmology (for a still very actual review see Bohm and Hiley book \cite{Hiley}).  Yet the precise story and chronology of the various steps leading Bohm to the redaction of his famous articles in 1952 is not exactly known since part of the correspondence between David Bohm, Louis de Broglie and also Wolfgang Pauli has been lost probably during the exile of Bohm to S\~{a}o Paulo in October 1951 due to his persecution by the House Un-American Activities Committee (HUAC) \footnote{We remind that Bohm left the United States for Brazil in October 1951 after refusing to testify against his purported links to the Communist Party to HUAC \cite{Peat}.  }. Therefore, the following chronology will be necessarily sketchy.  First, it is known that the story started with the writing by David Bohm of his undergraduate textbook \emph{Quantum theory} \cite{Bohm1951} which was finalized in 1951 and sent to major scientists such as Pauli and Einstein for approval. Importantly, the book contains a `proof that quantum theory is inconsistent with hidden variables' obtained in relation with  the Einstein Podolsky Rosen (EPR) paradox \cite{EPR} and the principle of complementarity of Bohr. Bohm therefore concluded on the nonexistence of `hidden variables underlying quantum mechanics'. Albert Einstein was very much interested by the book and gave a call to Bohm to discuss with him. The content of this discussion was summarized by Bohm himself in an interview made by his friend Maurice Wilkins in 1980 \cite{BohmWilkins1980}, the story is also recounted in Bohm's  biography written by David Peat \cite{Peat} (see also \cite{PeatBohm} and \cite{Bohm1982} which include other recollections of the whole story by Bohm, and finally Max Jammer book and article \cite{Jammer,Jammer2}). The main subject of the discussion concerned the EPR paradox \cite{EPR} which was analyzed in Bohm's Book. Einstein felt unconvinced by the `orthodox' analyzes made by Bohm who at that time still accepted the orthodox view taught by Niels Bohr and others concerning the general interpretation of quantum mechanics. In particular, Einstein's `\textit{objections were that the theory was conceptually incomplete, that this wave function was not a complete description of the reality and there was more to it than that}' \cite{BohmWilkins1980}. After the discussion Bohm changed his mind and started thinking about an alternative description of quantum mechanics which would preserve the classical credo of realism  and causality advocated by Einstein \cite{BohmWilkins1980}. The results of his work was a version of the pilot-wave theory already discovered by de Broglie in 1927. However, at that time Bohm apparently ignored the priority of de Broglie. In a letter to astrophysicist \'{E}vry Schatzman  (who was in touch with Jean Pierre Vigier and de Broglie) Bohm summarized: 
 \begin{quote}`\textit{Finally, I decided for a causal interpretation within few weeks, I hit upon the idea which I published not knowing about de Broglie's work until later}' \cite{BohmSchatzmanSeptember1952}.\end{quote} 
 In the same letter Bohm also confirmed the importance of the EPR paradox in the whole affair. \\
 \indent  We now remind that in 1925 after completing his doctorate thesis de Broglie \cite{debroglie1925} sought deeper for a mechanical deterministic model able to explain the wave-particle dualism of quantum matter. He then proposed an approach known as the `double-solution program' in which the quantum particle is a point-like object generating an extended $u$-wave which subsequently guides and affects its motion during propagation around obstacles and external potentials \cite{debroglie1926,debroglie1927}. The model was mathematically too ambitious at that time (the issue is still unsolved nowadays, for recent reviews see \cite{Fargue,Durt}). Therefore, in 1927 at the 5$^{th}$ Solvay Congress in Brussels \cite{Brussels,Valentini,debroglie1930} he instead presented the simpler pilot-wave model in which the point-like particles are guided by the $\Psi$-wave solution of Schrodinger's equation in the many-particle configuration space (see also Kennard contribution \cite{Kennard1928}).  However, due to several technical reasons (some of them being discussed later) de Broglie abandoned his theory after 1928-30. He went back to it (i.e., within the context of the double-solution theory) only 25 years later after the publication of Bohm's articles \cite{Bohm1952a,Bohm1952b} and the strong implication of Vigier in the collaboration to the double-solution program. Moreover, we stress that de Broglie was already prepared to change his mind as visible for instance in the evolution of his university-lecture notes during the period 1950-1951 which were later published in a book as a historical legacy \cite{debroglieHeis}. Interestingly this book contains some early negative reactions to a version of Bohm manuscript sent to de Broglie during the summer 1951 and a brief criticism of the pilot-wave theory \footnote{The title of Bohm's manuscript indicated by de Broglie in \cite{debroglieHeis} was: `A suggested new interpretation of the quantum theory'. This title corresponds  neither with ACCIQT nor to the final title \cite{Bohm1952a,Bohm1952b}. We don't have any explanation for these differences. It could be that Bohm finally sent a longer version of the manuscript different from the published version or simply that de Broglie mixed the titles. However, it is interesting to see the word `new' in the title which emphasizes the need for Bohm to insist on the novelty of his own contribution.  }.              \\
\indent Back in 1951, Bohm's wrote a draft summarizing his ideas and during the summer sent a version to several known physicists including Einstein, Pauli and de Broglie \cite{BohmWilkins1980}. Following \cite{PeatBohm} the draft contained only a treatment of the single particle case. As Bohm wrote his preliminary theory was complete 
\begin{quote}`\textit{at least in a one particle system which is as far as I got at that time}' \cite{PeatBohm}, p.36. 
\end{quote} 
Pauli answered very quickly criticizing the full hidden variable approach and commenting to Bohm that:
 \begin{quote}`\textit{It was old nonsense that de Broglie had done in 1927. They'd had the Solvay Congress there and that he had demolished de Broglie there}' \cite{BohmWilkins1980}.
 \end{quote}
 One of the famous objection given by Pauli concerned the interaction between an incident particle non-elastically scattered by a Fermi rotator. De Broglie had already sketched the good answer in 1927 \cite{Brussels,Valentini} based on the limited extension of wavepackets in the configuration space \footnote{The answer of de Broglie in 1927 was already close from the one given by Bohm in 1952. In 1927 de Broglie answered: 
`\textit{The whole question is to know if one has the right to assume the wave $\Psi$ to be limited laterally in configuration space. If one has this right, the velocity of the representation point of the system will have a constant value, and will correspond to a stationary state of the rotator, as soon as the waves diffracted by the $\phi$-axis will have separated from the incident beam}'  \cite{Brussels,Valentini}.
 In a letter to Pauli from July 1951 Bohm wrote:  `\textit{If you had chosen an incident wave packet, instead, then after the collision is over, the electron ends up in of of the outgoing wave packets, so that  a stationary state is once more obtained.}' (letter 1263 in \cite{BohmPauli1951}). } But in his letter to Bohm Pauli probably presented it as a fatal objection to de Broglie `\textit{which sort of really knocked him out}' \cite{BohmWilkins1980}.  De Broglie also answered to Bohm \footnote{In \cite{PeatBohm,Peat} it is precised that the de Broglie replied to Bohm before Pauli while this is not so clear in \cite{BohmWilkins1980}.} \cite{BohmWilkins1980}  explaining that he already developed the idea in 1927 and abandoned it in part due to Pauli's objections and mostly for others more fundamental reasons related to the measurement theory and the notion of wave-function collapse in quantum mechanics. The objections of de Broglie (recollected in  \cite{debroglieHeis,deBroglie1953,deBroglie1956}) are interesting and are already given in his 1930 review of the pilot-wave theory \cite{debroglie1930}. First, there is the question concerning the interpretation of `empty branches' of the $\Psi$-wave after scattering processes. Indeed, in the usual interpretation these empty waves collapse and have no subsequent physical effect. De Broglie wrote: \begin{quote}`\textit{it is difficult to conclude otherwhise that the wave is not  a physical phenomenon in the old sense of the word}' \cite{debroglie1930}.
 \end{quote} We stress that de Broglie's reasoning is a direct consequence of a discussion made by Einstein at Brussels in 1927 when he objected that: 
\begin{quote}`\textit{But the interpretation, according to which $|\Psi|^2$ expresses the probability that this particle is found at a given point, assumes an entirely peculiar mechanism of action at a distance, which prevents the wave continuously distributed in space from producing an action in two places on the screen}' \cite{Valentini}, p. 441.
\end{quote} Here, is clearly seen an issue with non-locality which rebutted both Einstein and de Broglie. As a second reason given by de Broglie for criticizing the pilot-wave theory, there is the problem of the energy conservation during scattering since in general the quantum potential responsible for the quantum effects  is time dependent. The source of this quantum energy is thus questionable. However, de Broglie emphasized \cite{debroglie1930} that as soon as the scattering ends up one will find one and only one of the energy allowed  by the standard quantum interpretation. Still, he doubted about this picture and wrote: \begin{quote} `\textit{herein lies one of the essential differences between the pilot-wave theory and the point of view of Bohr and Heisenberg}' \cite{debroglie1930}.\end{quote} In the 1950's \cite{debroglieHeis,deBroglie1953,deBroglie1956} he added that the EPR paradox is not completely solved within the pilot-wave framework because the collapse of the empty branches corresponding to alternatives which are not realized assume nonlocality which de Broglie, like Einstein, abhorred as mentioned before. It is probable that de Broglie shared some of his doubts and frustrations with Bohm in his letter (as revealed in the Bohm-Pauli correspondence \cite{BohmPauli1951,PauliBohm1951}). After receiving these letters of de Broglie and Pauli Bohm talked once again to Einstein \cite{BohmWilkins1980} who also confirmed de Broglie priority: Einstein like Pauli was present at the 1927 Solvay Congress and he was one of the few, with Brillouin, not to oppose directly to the pilot-wave interpretation contrarily to Pauli who gave detailed objections  \cite{Brussels,Valentini}. Einstein had his own objection to Bohm's theory also discussed in a  slightly different form at the Solvay Congress: two letters of Bohm to Einstein of 1951 mention this issue \cite{BohmEinstein1951} which concerned the interpretation of stationary states for the single-electron motion in a one dimensional box.  Indeed, Bohm's theory implies a single-particle resting in the box with null velocity $v=\nabla S/m=0$ and this even for high energy levels in mere contradiction with the classical intuition and the correspondence principle. This issue became the subject of section 5  in \cite{Bohm1952b}  (mirroring the two Bohm's letters of 1951 \cite{BohmEinstein1951}) and was subsequently discussed by Einstein, Bohm and de Broglie in \cite{Born} at the invitation of Einstein.\\
\indent  From that moment we know that Bohm strategy was to complete his first draft in order to reply to the objections made by Pauli at Brussels (and subsequently by de Broglie in a book published in 1930 \cite{debroglie1930}) and therefore to carry some new elements to the discussion for the many-body problem   \cite{BohmWilkins1980,PeatBohm}. This was clearly one of the main subject of the Bohm-Pauli correspondence during this period \cite{PauliBohm1951,BohmPauli1951}. The second issue focused on the paternity of the  pilot-wave theory. The final version in particular contained a theory of quantum measurements for the many-particle system which allowed Bohm to answer most of the questions raised by Pauli and de Broglie.  We further know from a reply of Pauli written in December 1951 that  Pauli urged Bohm to recognize the priority of de Broglie concerning the pilot-wave theory \cite{PauliBohm1951}. Pauli wrote: \begin{quote}`\textit{It should be also stated that de Broglie had formulated already the quantum-potential energy }' \cite{PauliBohm1951}.\end{quote} This letter was one of the last of a long correspondence between Pauli and Bohm concerning this manuscript. We stress that while several of Bohm's letters were recovered \cite{BohmPauli1951} only one of Pauli survived  and the annotated manuscript by Pauli was lost.  While Bohm resisted for a while to admit it he finally acknowledged the priority of de Broglie \footnote{In a letter to Pauli from November 1951 Bohm already acknowledged this point. He wrote `\textit{I have changed the introduction to give due credit to de Broglie, and have stated that he gave up the theory too soon (as suggested in your letter).}' (letter 1309 in \cite{BohmPauli1951} from  November 1951).} but only after emphasizing that: \begin{quote}`\textit{If one man finds a diamond and then throws it away because he falsely concludes that it is a valueless stone, and if this stone is later found by another man who recognize its true value, would you not say that the stone belongs to the second man?}' Letter 1290 from October 1951 in \cite{BohmPauli1951}.  \end{quote} 
It is interesting to note that the same `diamond' parabolic saying was also sent by Bohm to Schatzman in a undated letter of 1951 \cite{BohmSchatzman1951,Besson} (we can however deduce that the letter was sent in September 1951 since Bohm mentioned his future travel in Brazil, i.e., in October 1951, and also commented a recent paper of de Broglie written in the beginning of September 1951 \cite{debroglieSept1951}). In the same letter Bohm wrote to Schatzman that his manuscript `A suggested interpretation...' is going to be published in the Physical Review the 15$^{th}$ of December 1951. This date is very interesting  for two reasons. First, because the paper was  not actually published in 1951 but in January 1952 probably due to editing  procedures. Indeed, Bohm  without waiting submitted his longer version  to Physical Review in July 1951 while he continued the correspondence with Pauli until 1952. The subsequent corrections provided by Bohm after many discussions with Pauli probably delayed the publication until January 1952.  The second and most important reason why this date is interesting is that in ACCIQT Bohm also mentioned a longer forthcoming paper (announced as reference 2 in ACCIQT) to be published in the Physical Review  the 15$^{th}$ of December. The year is missing  but the comparison with \cite{BohmSchatzman1951,Besson} allows us to deduce that it was in 1951. From this we can deduce that i) ACCIQT was mainly intended to be a short summary of Bohm ideas already submitted to Physical Review, and ii) that ACCIQT was indeed sent to de Broglie before December 1951. Moreover, from the S\~{a}o Paulo address on top of ACCIQT we can also deduce that the manuscript was necessarily sent after the beginning of August 1951 since we know \cite{PeatBohm} that Bohm's contract at the University in Brazil was fixed (thanks to recommendations of Einstein and Oppenheimer) only after that date. Importantly, at the end of ACCIQT Bohm (in a note `added after the work reported here had been completed') gives full references  to previous publications by de Broglie and Madelung \cite{Madelung1926} (but not by Rosen \cite{Rosen1945} which will be added in the final version \cite{Bohm1952a,Bohm1952b}). This added notes result probably from the interaction between Bohm and Pauli and Bohm and Einstein.   It should be emphasized that ACCIQT starts with a single-particle analysis like in the first version sent to Pauli and de Broglie (as explained by Bohm himself \cite{PeatBohm}). This shows that the ACCIQT draft was combining some elements of the first paper with new ideas to be developed in a subsequent manuscript \footnote{It should be emphasized that there is no annotation of de Broglie on the ACCIQT draft contrarily to other manuscripts of Bohm found in the de Broglie Archives. This could suggest that de Broglie already knew the main content of the paper ACCIQT (i.e., already discussed in a previous draft).}. Moreover, Bohm underlined some words in order to emphasize their importance (we go back to this issue in Section \ref{section3}).  Similar words are also employed in the letter of Bohm to Schatzman \cite{BohmSchatzman1951,Besson} and they clearly refer to answers of criticisms made by Pauli and de Broglie \footnote{To make the situation even more complicated Bohm joined to the Schatzman letter a mimeographed copy of his new longer manuscript for Vigier and his collaborator R\'{e}gner. This manuscript had already the final title `A suggested...' but was not yet the final version submitted to the physical review as recognized by Bohm himself \cite{BohmSchatzman1951,Besson}.}. \\
\indent We now go back to the chronology issue. From our inquiry we thus know that the paper ACCIQT sent to de Broglie at least after the start of  August was written after Bohm's letter of July 1951 (letter to Pauli 1263  from July 1952 in \cite{BohmPauli1951}) where Bohm already acknowledged having sent to Pauli a much longer and complete version of his work, i.e.,  answering all the fatal objections raised by de Broglie and Pauli. However, the two central questions that we should ask now are : \\ 
\indent i) Why did Bohm sent the shorter version ACCIQT to de Broglie in August whereas he had already finished a better and longer work that he submitted to the Physical Review and sent to Pauli in July? We believe that the response to this question is related to the difficult and ambiguous relation existing between Bohm and de Broglie during this period.\\ 
\indent ii) Where did Bohm submit this work? We will try to propose answers to this difficult question afterward. \\
\indent Clearly, to answer the first question it is important to remind that the interactions between Bohm and de Broglie at that time were complicated by this priority issue. In \cite{BohmSchatzman1951} Bohm comments that: 
\begin{quote} `\textit{I have heard also that de Broglie has  recently published an article on the subject, which I have no yet time to read. I recently received a letter from de Broglie in which he took great pains to claim credit for the ideas. My answer to him was to admit that he suggested the method in 1926, but to point out that because he did not carry it to its logical conclusion he came to the erroneous conclusion that the idea does not work}' \cite{BohmSchatzman1951,Besson}.\end{quote}  The paper of de Broglie was the note to the Academy of Science of September 1951  \cite{debroglieSept1951} in which de Broglie after thanking Bohm for sending him a draft of his work subsequently recalled his unarguable priority and at the same time criticized the pilot-wave theory for being too abstract to be realistic. 
In a English edition of his book `Physics and Microphysics' de Broglie added that he just received the work of Bohm and that he still found the pilot-wave theory too crude and abstract since:
\begin{quote} `\textit{Reflecting de novo on these problems, it always appears to me that this last form of my ideas of 1927 is impossible to accept; if we return to the objectivistic point of view of classical physics, we could not, in effect, admit that a corpuscle would be guided in its movement by a $\Psi$-wave of wave mechanics since this $\Psi$-wave is only a representation of probabilities.}' \cite{debroglie1955}.\end{quote} De Broglie subsequently explained that the double-solution theory, i.e., with $u$-waves propagating in the 3D space, is more physical and therefore closer to a realistic and causal description of nature than the pilot-wave model involving `fictitious' $\Psi$-fields. The approach followed by de Broglie was certainly closer in spirit to the one made by Einstein and this clearly appeared in subsequent works done together with Vigier where the analogy with general relativity played a fundamental role in the attempts to develop the double solution program in the 1950's (for recollections concerning these issues see \cite{deBroglie1953,deBroglie1956,Vigier1956}). \\
\indent Therefore, we see  that  the comments of de Broglie in his note of September were not only a question of priority with Bohm but were directed against himself and his own pilot wave-theory of 1927. Only few months later de Broglie went back to the double-solution with Vigier but still criticizing the pilot wave model for being insufficient. Moreover, in a letter to L\'{e}on Rosenfeld dating of December 1951 \cite{DestouchesRosenfeld1951,Besson} Jean Louis Destouches wrote that de Broglie had priority quarrels with Bohm: De Broglie first reminding Bohm his priority, Bohm then promising to give references to de Broglie works, and finally Vigier and also Tonnelat motivating de Broglie's  change of mind for returning to the double-solution realistic theory. Bohm became particularly defiant concerning de Broglie. This is clearly visible in a letter of Bohm to Schatzman written in July 1952 where at the same time Bohm wished a collaboration with Vigier and R\'{e}gnier but wanted to hide information to de Broglie \cite{BohmSchatzman1952,Besson}. This communication issue was solved progressively during the 1950's \footnote{As other signs of an improving relation between Bohm and de Broglie we can mention that de Broglie wrote a very positive preface for Bohm's 1957 book `Causality and chance  in modern physics' \cite{Bohm1957}. Moreover, when Rosenfeld wrote a vitriolic review for this book in the journal Nature of  March 1958 \cite{Rosenfeld1958} de Broglie replied defending Bohm \cite{deBroglie1958} (see also Rosenfeld's short reply \cite{Rosenfeld1958b}). }  when both parties saw the advantage of a fruitful collaboration (specially with the Bohm-Vigier work). From this, we can speculate that Bohm's manuscript ACCIQT sent in August 1951 was hiding some results obtained during the discussion with Pauli. On the defense of Bohm it should be emphasized that his situation was particularly complicated and difficult at that time. The consequences of his exile from US on his professional and private life were certainly a source of stress and anxiety affecting his relationship not only with those like  Pauli, Rosenfeld, Heisenberg which were strongly critical and skeptical but also  with potential allies or competitors like de Broglie having the ability to disseminate results very quickly in French journals like the Comptes Rendus \footnote{This fear is enlighten by Bohm in the letter he sent to Schatzman in 1952 \cite{BohmSchatzman1952,Besson}.}. Again, the situation progressively changed  when it became clear to Bohm that de Broglie mainly wished to advocate and develop his double solution theory which was thus a different program not necessarily conflicting with Bohm own hidden-variable approach. De Broglie certainly understood the advantage to give support to Bohm and Vigier works for the future development of quantum mechanics. In a letter sent to Bohm in February 1952 de Broglie wrote 
\begin{quote}`\textit{I have heard that you could perhaps come to France this year. I would be very happy  of this because I believe that we could collaborate usefully and that direct discussions between us would be useful.}' \cite{deBroglieBohm1952}.
\end{quote} Bohm recognized this change in de Broglie reaction and  in a letter to his friend Melba Phillips in early 1952 Bohm wrote: 
\begin{quote} `\textit{De Broglie is now fairly friendly to me, saying in a letter tha I have carried the pilot-wave theory much further that he did in 1927}' \cite{BohmMelba1952}. \end{quote}
\indent Now, we would like to understand where the manuscript ACCIQT was submitted or intended to be submitted. It is indeed very improbable that  Bohm wrote an article only for being read by de Broglie. While we don't have any clear and definitive answer to this question we would like to speculate two possible answers.   The first possible answer came to us after reading the biography  of David Peat where it is mentioned \cite{Peat} p. 128 that in a letter to his close friend the mathematician Miriam Yevick (written in January 1952) Bohm answered to de Broglie critical article of September 1951  \cite{debroglieSept1951}. Bohm wrote: 
\begin{quote} `\textit{I have answered de Broglie's criticisms (published in Comptes Rendus) of my article by sending a letter to the Phys. Rev., which should come out in a few months. De Broglie apparently didn't read my article, but simply re-iterated Pauli's criticisms, which led him to abandon the theory, but did not point out my conclusion that these objections are not valid. }' \cite{BohmYevick1952}.\end{quote} 
However, no paper concerning de Broglie never appeared in the Physical Review concerning this issue \footnote{Peat misleadingly gave a reference to a different published article \cite{Peat} p. 128.}.  Actually, the reply to de Broglie critical analysis was given in the second 1952 article \cite{Bohm1952b} so that the need of a specific letter disappeared.
 Yet, there is a second possibility for explaining ACCIQT. Indeed, in a previous letter written in November 1951 to his friend Yavick Bohm explained that: 
\begin{quote} 
 `\textit{Also, I sent a brief article to Massey with the suggestion that he publish it in Nature, and telling him that I hope to visit England this June.}' \cite{BohmYevickNovember1951}. \end{quote} 
Sir Harrie Stewart Wilson Massey  was a Australian physicist specialist in nuclear physics during the second world-war, and one of his project involved Bohm  (see \cite{Peat}, p.65). Therefore, good contacts for publishing ACCIQT were potentially possible for Bohm  in England where Massey worked as head of the University College London, Physics Department since 1950.
Interestingly, in a subsequent letter written in December 1951 to Yavick Bohm wrote: 
\begin{quote} 
`\textit{I have answered Pauli and am awaiting an answer from him, and am still awaiting an answer even from de Broglie (Thus far have I sunk!) Incidentally, I sent a manuscript on quantum theory almost a month ago to Massey in England, with the suggestion that he have it published in Nature. I still haven't heard from him yet. I have a suspicion that he is losing his nerve, and is both afraid to send it in and afraid to tell me that he is not sending it in. But in a few weeks I shall know (That is, if I am not kidnapped and spirited back to the good old USA as that fellow in Holland was)}' \cite{BohmYevickDecember1951}. 
\end{quote}  
Here we have different useful information: On the one side, we see that Bohm was still corresponding with  Pauli and de Broglie in the late 1951. On the other side, we have the word `incidentally' used by Bohm in his letter to express the fact that the manuscript sent to Massey played a subordinate role in this story. Again all this tends to confirm the idea that the manuscript sent to de Broglie was intended to have several roles: i) on the one hand, it was written to produce a reaction on de Broglie side and ii) on the second hand, this would allow Bohm to publish in the Journal Nature a kind of `teaser' for his future work to appear in 1952 \cite{Bohm1952a,Bohm1952b}. Unfortunately, the paper never appeared in Nature and this is perhaps related to the ambiguous reception by Massey. This is as far as we were able to go concerning the chronology and history of ACCIQT. In the next section  we will consider and analyze more in details the physical content of this manuscript.    
  
\section{Contents of the manuscript}
\label{section3}
\indent The manuscript ACCIQT starts with a condensed description of the pilot-wave theory for a single electron. The Newtonian (i.e., second-order) law of motion for a quantum point-like electron reads 
\begin{eqnarray}
m\frac{d^2}{dt^2}\textbf{x}(t)=-\boldsymbol{\nabla}[V(\textbf{x}(t),t)+U(\textbf{x}(t),t)]\nonumber
\end{eqnarray} where $U(\textbf{x},t)=\frac{-\hbar^2}{2m}\frac{\boldsymbol{\nabla}^2|\Psi(\textbf{x},t)|}{|\Psi(\textbf{x},t)|}$ is the famous quantum potential. Bohm's stresses the role of the additional assumption $\mathbf{p}=m\frac{d}{dt}\textbf{x}=\boldsymbol{\nabla}S(\textbf{x}(t),t)$  which will  be more detailed in the longer work in preparation cited as reference 2 in the manuscript ACCIQT.\\ \indent  Moreover, Bohm emphasizes in several places in ACCIQT that the new theory `\textit{leads to precisely the same results for all experiments as are given by the usual interpretation}'. This was clearly a point of disagreement with Pauli and de Broglie at that time and the same response was enlighten in the letter of Bohm to Schatsman \cite{BohmSchatzman1951,Besson} concerning the longer manuscript announced in December 1951. The fact that the new theory is empirically equivalent to the orthodox one relies on a description of many-body interactions like for the photo-electric effect which is mentioned in the manuscript. While this is very sketchy in ACCIQT more detailed analysis is promised by Bohm in the longer work announced for December 1951 (i.e., as reference 2 in ACCIQT).  It is remarkable that no single word is devoted to Pauli's famous objection made during the Solvay Congress and concerning the Fermi rotator interacting with a second particle. Neither the manuscript discusses the EPR paradox which will be analyzed in \cite{Bohm1952b}. In place there is a discussion of the famous von Neumman no-go theorem which conflicts with the mere existence of Bohm's theory. Bohm affirms that in his theory hidden variables must also be assigned to measurement apparatus and that this provides a loophole in von Neumann's proof. This naturally means  that macroscopic systems such as apparatus and observers are described by the same dynamics and theory as the observed systems. In other words, the qualitative separation or `shift' between observed and observers which lies at the core of the orthodox interpretation vanishes completely in the hidden-variable theory presented by Bohm.  A similar claim was developped in the second 1952 published article \cite{Bohm1952b}. We now know that the von Neumann's proof relies on some unwarranted additivity assumptions which were clearly identified and debunked by John Bell in 1966 \cite{Bell1966} (and before that by Grete Hermann in 1935 \cite{Hermann1935}). The analysis of Bohm can therefore only have a qualitative value. It should be emphasized that de Broglie also published his own criticism of von Neumann's theorem in his note of September 1951 \cite{debroglieSept1951}. One very interesting point raised by de Broglie against von Neumann's theorem is a proof by contradiction (reductio ad absurdum): the  mere existence of the pilot-wave theory shows that  von Neumann's proof is generally wrong.   De Broglie also included a discussion of quantum measurements. Following his discussion von Neumann's theorem fails because the different statistics associated with measurements of complementary observables such as position $x$ and momentum $p$ can not be obtained in the same experiment whereas in the pilot-wave theory the actual position of the particle $x(t)$ plays a fundamental role imposing  the probability of presence $|\Psi(x)|^2dx$ as more fundamental and valid at any time (de Broglie analysis is further developed in \cite{debroglieHeis,debroglie1957}). Likewise, the analysis of de Broglie (like the one made by Bohm) is very general and doesn't identify the specific mathematical issue invalidating the generality of von Neumann's proof.  In retrospect, when we re-analyze de Broglie and Bohm criticisms of von Neumann's proof with the results obtained by Hermann and Bell \cite{Hermann1935,Bell1966} we see that indeed the pilot-wave theory contradicts the additivity assumptions made by von Neumann.\\
\indent In the same context, ACCIQT contains a still sketchy argumentation concerning the measurement theory which will be extended in the final papers of 1952. This measurement theory is generally considered to be a key contribution of Bohm in \cite{Bohm1952a,Bohm1952b}. We stress that quantum measurements were already analyzed by Bohm in \cite{Bohm1951} from a Bohrian orthodox perspective. Therefore, Bohm was well designated to develop a measurement theory for the pilot-wave interpretation. From Bohm's interview  \cite{BohmWilkins1980} we know that it is only after discussing with Einstein that Bohm decided to discuss measurement processes in response to Pauli's objections. Again, this suggests that the manuscript ACCIQT was already a modified and hybrid version of the original one sent to Pauli, Einstein and de Broglie. Generally speaking, Bohm tried to emphasize the novelty and superiority of his work in ACCIQT compared to de Broglie's study by underlining some words such as `all, individual, practical, every...' playing a key role in the message of the paper. For example the opposition between `individual' and `intrinsically statistical' emphasizes the novelty of the causal interpretation with respect to the orthodox interpretation limited to statistical ensembles and neglecting individual systems. It is important to stress that this dilemma was at the core of Einstein criticisms of the orthodox quantum interpretation. For example in his popular book written with Leopold Infeld the comparison is made between quantum statistical mechanics and statistics in classical physics. Here they wrote : 
\begin{quote} 
`\textit{But in quantum physics the state of affairs is entirely different. Here the statistical laws are given immediately. The individual laws are discarded}' \cite{EinsteinInfeld}. \end{quote} 
\indent It should be emphasized that in a letter written in October 1951 Bohm commented some important changes made in the much longer version sent to Pauli probably before July \footnote{The date can be inferred from a letter to Pauli written in July (see letter 1263 in \cite{BohmPauli1951}) where Bohm described the content of his new and much longer manuscript now separated in two parts I and II. We also remind that the two papers were submitted to the Physical Review the 5 July 1951.} (see letter 1290 in \cite{BohmPauli1951}). One of these changes concerns discussions of a `molecular chaos' hypothesis whose importance was weakened in the version sent to Pauli and in the final articles of 1952. In the letter to Pauli Bohm explained that he doesn't `\textit{need to use molecular chaos}' \cite{BohmPauli1951} to justify Born's rule $P(x)=|\Psi(x)|^2$. Indeed, as shown by Bohm  the Born rule holding true at one time will be so at any other time (this is done in analogy with Liouville theorem's in classical mechanics). This idea was already known by de Broglie in 1926 who however tacitly assumed the equality $P(x)=|\Psi(x)|^2$. Still, this molecular chaos hypothesis is clearly discussed in ACCIQT as a key ingredient in the theory (even though not necessary).  More precisely, in ACCIQT Bohm mentions (in connections with molecular chaos) the possibility that Born's rule $P(x)=|\Psi(x)|^2$ relies on the chaotic Brownian motions of particles interacting with other systems.  Moreover, in ACCIQT Bohm affirms that `\textit{it can be shown that after sufficient interactions have taken place the probability density will tend to approach} $P(x)=|\Psi(x)|^2$' (in the second 1952 article Bohm \cite{Bohm1952b} mentions the existence of a H-theorem reminiscent of Boltzmann's theory for justifying the tendency to reach statistical equilibrium). We can also point out that in ACCIQT a too fast reading could have easily led Pauli to believe that the diffusion and relaxation  process linked to molecular chaos are essential and not contingent to the theory.  Bohm's explanation in his letter to Pauli was intended to be clarifications of these misunderstandings. In his October letter Bohm  wrote :
\begin{quote} `\textit{Thus the complicated and chaotic motion that occurs in interaction, to which you and de Broglie objected so strongly, is just what is needed to establish the ensemble} $P=|\Psi|^2$. \textit{It is therefore something that is desirable in the theory' }(letter 1290 in \cite{BohmPauli1951}).\end{quote}  The final version of 1952 clarified this misunderstanding  by emphasizing more cautiously what is central and what is not. Once again, this discussion strongly suggests that ACCIQT was a still primitive version of the 1952 paper very similar to the original manuscript. Furthermore we remind  that all these important ideas of Bohm were actually further developed in papers written in 1953 \cite{Bohm1953}, and in 1954 conjointly with Vigier \cite{Bohm1954}, and in a book published in 1957 \cite{Bohm1957} but we saw that these ideas were already discussed and partly understood in 1951. Interestingly, Pauli came back to the issue about probabilities in an article written in honour of de Broglie in 1953 \cite{Pauli1953} where he again claimed that Born's rule is not justified in the pilot-wave theory (interestingly Pauli didn't question de Broglie directly but focused his critical work on Bohm). \\
\indent A final ingredient of  ACCIQT was to suggest some empirical differences between his theory and the standard approach. Bohm's suggested new features occurring at physical scales below $10^{-13}$ cm where non-linearities could modify the law of motions. Similar suggestions were developed in all Bohm's subsequent works.  This point is indeed crucial for Bohm since a difference with standard quantum mechanics would allow him to reply to criticisms made by those like Pauli and Heisenberg claiming that the theory is purely metaphysical. Pauli for instance  called this hypothesis  `\textit{a check which can not cashed}' \cite{PauliBohm1951} whereas Heisenberg ironically quoting N. Bohr compared Bohm's program to the `\textit{hope that it will later turn out that sometimes $2\times2=5$ for this would be of great advantage for our finance}' \cite{Heisenberg}.  In particular the uncertainty principle mentioned in ACCIQT could lose his general validity. This issue became the core of the collaboration in the 1950's and 1960's between Bohm, Vigier, and de Broglie who all accepted the idea of a complex `sub-quantum dynamics' strongly  affecting the microscopic quantum evolution \cite{Bohm1952a,Bohm1952b,Bohm1953,Bohm1954,Vigier1956,deBroglie1956}.  De Broglie expected very much from the Bohm-Vigier work and was delighted by the new possibilities. In the preface to Bohm's book he wrote:  \begin{quote}
`\textit{One can it seems to me, hope that these efforts will be fruitful and will help to rescue quantum physics from the cul-de-sac where it is at the moment}' \cite{Bohm1957}.
\end{quote}  
Clearly, the ACCIQT manuscript of 1951 is a precious one since it reveals some of the early fundamental ideas and concepts of Bohm to be developed  during the rest of his life.  
\section{Conclusion: Bohm and the implicate order }
\label{section4}
\indent As we saw, the manuscript ACCIQT contains already many of the key elements  to be discussed by Bohm in his subsequent works. Most importantly, the draft discusses the role of statistics and stochasticity (i.e., how to justify   Born's rule) and the possibility of new physics in the sub-nuclear regime below  10$^{-13}$ cm, where present theories do not seem to be adequate (probably nowadays we would better consider the Planck length 10$^{-33}$ cm as a typical scale where gravitation is going to play a critical role in quantum physics). The ideas of Bohm in ACCIQT also included the role of man- body interactions  which are only briefly sketched (in particular in relation with the von Neumann no-go theorem).   Later, development emphasized the role of non-locality which is already discussed in \cite{Bohm1952b} in connection with the EPR paradox.  \\
  \indent One of the key element for Bohm during the 1950's  was the role played by Marxism and by the book \emph{Materialism and Empirio-criticism} written by Lenin in 1908 \cite{Lenin}. In this book about dialectical materialism Lenin strongly criticized the positivism of Ernst Mach and strongly supported causality for a clear description of matter in space and time.  The reject of positivism by Bohm is already very clear in the manuscript ACCIQT where the word causal appears in the title. The criticism of Mach and positivism was more detailed in the 1952 articles  \cite{Bohm1952a,Bohm1952b} and in his book \emph{Causality and Chance in Modern Physics} \cite{Bohm1957}.  In that book Bohm described his approach of an infinity of levels where he conceived mechanism  and strict determinism a la Laplace as only a first approximation for a causal theory.  Bohm saw the presence of stochastic elements as a sign that the theory of the Universe must be scale dependent (a topic which also played a role in his research in condensed matter physics, i.e., related to the development of the theory of renormalization in connection with collective quantum excitation in plasma:  the so called plasmons). In his frame work, which  was also advocated by his Marxist friends Vigier and Schatzman (see \cite{Besson,Peat} for a clear discussion), the presence of stochastic elements corresponded to an approximation for a causal and realistic underlying physics, i.e., at the sub-quantum level. Yet, for Bohm it became clear that at such lower scales and higher energies the structure of physical laws  should not re-establish the strict determinism of classical physics. Instead, Bohm conceived the existence of a new emerging set of laws applicable to the lower spatial scales and still containing random and stochastic elements mixed together with deterministic rules. This is a bit like in Langevin's description of Brownian motion where a random and fluctuating force $\eta$ is added to the classical force $F$ in the Newtonian dynamics $m\ddot{x}=F+\eta$. This process was for Bohm infinite, meaning that new laws and properties emerged at different scales (the idea generalized some early speculations made by Lenin about `the inexhaustible electron' \cite{Lenin,Bunge1950}). Moreover, for Bohm, additionally to stochastic elements, the presence of a quantum potential $U=\frac{-\hbar^2}{2m}\frac{\boldsymbol{\nabla}^2|\Psi|}{|\Psi|}$ was a completely new feature which emphasized collective and nonlocal properties without classical counterpart. Indeed, for Bohm this quantum potential was not completely mechanical in the sense that its intensity doesn't fall out with the distance and this is very different from classical force like Newtonian gravitation or electromagnetic forces which strongly decay with the distance between particles. Also, this strange potential only depends on the local form of the wave function  but not on its absolute amplitude (i.e., the multiplication of $|\Psi|$ by a constant doesn't change $U$). For Bohm, like in his work with plasmons, this non-local and collective feature  was a specificity of the new quantum order. \\
  \indent However, the early model proposed in the 1950's was progressively abandoned by Bohm in the 1960-70's because of too much arbitrariness in the choice for the new quantum dynamics (this was reminiscent of criticisms already made by Pauli, Einstein and de Broglie). Instead, Bohm emphasized that the mere existence of the pilot-wave model is \textit{at least } possible.  As Bohm wrote in 1962 \cite{Bohm1962} in a comment to Heisenberg's  1958 book \cite{Heisenberg}: 
  \begin{quote} `\textit{While trying to find a way to remedy the absence of `actuality function' he\footnote{i.e., the author David Bohm} developed a definite example of an alternative interpretation,  which permitted the quantum theory to be extended so as to include them in a logically consistent way}' \cite{Bohm1962}.\end{quote} In other words, the pilot-wave model Bohm developed in 1951-52 already shows the possibility for an alternative causal quantum dynamics, i.e.,  clearly violating the usual Copenhagen credo that no such a model is even conceivable (a point which was also stressed by de Broglie in 1951 \cite{debroglieSept1951}). This minimalist view was often accepted and used by proponents of the so-called `Bohmian mechanics'. For example Bell wrote:  
  \begin{quote} `\textit{Why is the pilot wave picture ignored in text books? Should it not be taught, not as the only way, but as an antidote to the prevailing complacency? To show that vagueness, subjectivity, and indeterminism, are not forced on us by experimental facts, but by deliberate theoretical choice?}' \cite{Bell1982}.\end{quote} 
	However, it is central to recognize that this choice of a minimalistic interpretation was for Bohm like for de Broglie only a temporary expedient. Contrarily to  widespread beliefs   Bohm was not a proponent of Bohmian mechanics:  His approach should better been named Bohmian non-mechanics since a return to strict determinism is not expected. In the same article of 1962 Bohm explained that in his new approach  trajectories should be better considered as having a fractal nature:  the notion of resolution being central for defining a dynamics (even the notion of continuty of trajectories was not central).  In 1962 he also presented a paper (reproduced as chapter 4 in \textit{Wholeness and the implicate order} \cite{BohmImpl}) where he presented an alternative ontological interpretation generalizing quantum mechanics  and hoping that new experimental facts could be generated at high energy in the subquantum regime. However, in 1979 he went back to his old quantum potential approach after the important numerical work made by his collaborator Basil Hiley with two students \cite{Hiley1979}. Together with Hiley Bohm wrote the influential textbook \textit{The undivided Universe} \cite{Hiley} where many of his remarkable ideas were synthetized. This included the pilot-wave theory but also some speculations about the undivided wholeness, implicate order, and the concept of active information which all played a fundamental role in his work over the years. In particular, active information characterizes the specificity of the quantum potential carrying information through space and time in a non-local and thus highly non-classical or mechanical way. It is interesting to see that the old manuscript ACCIQT of 1951 mirrors this ultimate work containing implicitly all the questions raised from the start.  
\section{Notes added in proof} 
After this work was completed we learned about the work made by the historian Olival Freire in 2005 \cite{Freire2005,Freire2019} where the existence of Bohm's manuscript published here is briefly mentionned.  The manuscript we found and analyzed here is very probably an article submitted to Nature by S. W. Massey in 1951 and refused after some critical comments made by L\'eon Rosenfeld to the editors. This therefore answers  our second question in section \ref{history} concerning the journal where ACCIQT was submitted and confirms our hypothesis made at the end of section \ref{history} concerning a potential submission to Nature.          
\section{Acknowledgments} 
The authors wish to thank the Foundation Louis de Broglie for very interesting discussions motivating the present work. We specially thank interactions with Christian Thomas de Pange, Philippe Fr\'{e}bault, Michel Karatchentzeff,  Daniel Fargue and Georges Lochak, as well as for their help and supports and for answering many historical questions. Authors also thank  the archives Louis de Broglie hosted at the Academies des Sciences Paris for allowing reproduction of the present document. We thank Basil Hiley for interesting discussions and for allowing us to reproduce Bohm's manuscript. We thank the Archives David Bohm at the Birkbeck University of London (and in particular Emma Illingworth) as well as the Albert Einstein Archives at the Hebrew University of Jerusalem for providing us some precious documents of the David Bohm and Albert Einstein Correspondence. We thank Alexandre Matzkin for providing us information about Olival Freire work. We finally thank Jean Bricmont and L\'{e}na Soler for the key role they played in making this discussion and collaboration possible and for interesting discussions.  

\end{document}